\documentclass{article}
\usepackage[utf8]{inputenc}
\usepackage[english]{babel}
\usepackage{amsmath}
\usepackage{amssymb}
\usepackage{dsfont}
\usepackage[margin=1.1in]{geometry}
\usepackage{esint}
\usepackage{graphicx}
\usepackage{subcaption}
\usepackage{float}
\usepackage{mathtools}
\usepackage{xcolor}
\usepackage{setspace}
\usepackage[shortlabels]{enumitem}
\usepackage{csquotes}
\numberwithin{equation}{section}
\usepackage{hyperref}
\hypersetup{
            colorlinks=true,
            linkcolor=blue,
            urlcolor=blue,
            citecolor=blue,
            linktocpage=true
            }
\usepackage{fancyhdr}
\usepackage[font=small,labelfont=bf,justification=justified,format=plain]{caption}
\usepackage{authblk}
\usepackage{cite}

\DeclareMathOperator{\im}{Im}
\newcommand{\of}[1]{\left(#1\right)}
\newcommand{\off}[1]{\left[#1\right]}

\newcommand{\vac}{|\Omega\rangle}
\newcommand{\tvac}{|\Omega_\beta\rangle}
\newcommand{\dd}{\mathrm{d}}

\spacing{1.25}

\begin{document}

\title{\textbf{Modular conjugation for the chiral fermion in multicomponent regions on the torus}}

\author[1]{Nicolás Abate\footnote{Email: 
 \href{mailto:nicolas.abate@ib.edu.ar}{nicolas.abate@ib.edu.ar}}}
\author[2,3]{Mateo Koifman\footnote{Email: \href{mailto:mkoifman@df.uba.ar}{mkoifman@df.uba.ar}}}

\affil[1]{\small Instituto Balseiro, Centro Atómico Bariloche 8400-S.C. de Bariloche, Río Negro, Argentina.\medskip}

\affil[2]{\small Universidad de Buenos Aires, Facultad de Ciencias Exactas y Naturales, Departamento de Física. Buenos Aires, Argentina.\medskip}

\affil[3]{CONICET - Universidad de Buenos Aires, Instituto de Física de Buenos Aires (IFIBA). Buenos Aires, Argentina.}

\date{}

\maketitle

\vspace{-1cm}

\begin{abstract}
    We continue the study of the Tomita-Takesaki modular conjugation for a massless Dirac field in a generic multicomponent region in $1+1$ spacetime dimensions. In this paper we focus on the computations for a thermal state on a circle, namely on the euclidean torus. By analytic continuation from the modular flow we arrive at an explicit expression for the modular conjugation in this scenario and derive its relevant limits. In contrast to the case of the vacuum on the line, this new result has a non-local behaviour even for connected regions. It also presents a novel contribution coming from the purification one has to introduce in order to deal with a mixed state: a term that maps the algebra of operators of the region to a copy of the global one, the so called ``second world'' algebra.
\end{abstract}

\tableofcontents

\newpage

\section{Introduction}

Over the last years the study of entanglement and information-theoretic aspects of quantum field theory (QFT) has become increasingly important. For instance we can mention the well-known renormalization group flow irreversibility theorems, which were proven in several contexts using the strong subadditivity of the entanglement entropy or the monotonicity of the relative entropy \cite{Casini07,Casini12,Casini17a,Casini17b,Casini23a}. There have also been applications in the proof of several energy inequalities \cite{Blanco13,Blanco18a}; and even to holography, for example in the derivation of the linearized Einstein equations in the bulk from entanglement properties of the boundary conformal theory \cite{Faulkner14,Lashkari14,Blanco18b,swingle14}.

In many of these applications, the knowledge of entanglement or modular Hamiltonians has proven to be extremely valuable. Although the first known examples of modular Hamiltonians where local \cite{bw75,bw76,Casini11,Cardy16}, in general they are very complicated non-local quantities \cite{Casini09,Arias18,Blanco19,DyG_19,Mintchev21}. Since it is defined as the logarithm of the density matrix, the modular Hamiltonian is ultimately ill-defined in any continuum theory such as QFT; but surprisingly, the dynamics it generates -- the modular flow -- does have a well-defined continuum limit. The modular flow emerges in a rigorous way in the context of the Tomita-Takesaki modular theory of operator algebras (see section \ref{section2} or \cite{summers05,witten18} for a review), which is a very rich topic concerned about the properties of the modular operator ($\Delta$) and the modular conjugation ($J$). It is the former operator which is related to the modular flow and hence the modular Hamiltonian.

Much less attention has been paid to the modular conjugation and explicit expressions are available in very few cases, most of them being vacuum states reduced to single component regions. Famous examples are the vacuum state of any theory reduced to the Rindler wedge (this is the well-known Bisognano-Wichmann theorem \cite{bw75,bw76}) or a conformal field theory (CFT) in a double cone region \cite{hisloplongo82}. More recently, an explicit expression for $J$ was found in the case of the vacuum state for a free massless fermion in $1+1$ spacetime dimensions (a chiral fermion) reduced to an arbitrary multicomponent region \cite{nuestro_paper}. Also, the modular conjugation for thermal states is known for the chiral fermion reduced to a double cone \cite{tonni22}. In this article we seek to expand this result to more general regions.

The paper is organized as follows. In section \ref{section2} we introduce the Tomita-Takesaki theory and discuss some of its properties and relation to QFT. In section \ref{section3} we describe the relevant aspects about the model we are interested in, a thermal state of the chiral fermion in $1+1$ spacetime dimensions living on a circle. Section \ref{section4} contains the main result of our work: an explicit expression for the modular conjugation operator for multicomponent regions in our model. Previously, we introduce the method employed to derive this result reviewing the recently found case of a vacuum state on the infinite line \cite{nuestro_paper}, which is simpler. Our computation shows some interesting novelties related to the non-zero temperature of the state, in particular it displays an explicit contribution form the copy of the theory (the ``second world'') one introduces to purify the state, and in this section we describe this contribution in detail. Also in this section we derive some interesting limits of our result and compare them with previously known examples, when possible. Finally, in section \ref{section5} we conclude with a discussion of our results. 

\section{Tomita-Takesaki theory}\label{section2}

Tomita-Takesaki theory is the study of von Neumann algebras admitting a cyclic and separating vector. We refer the reader to our previous work on the subject for a review of these concepts \cite{nuestro_paper}. The starting point is the introduction of the Tomita operator, defined over the dense set of states generated by such an algebra $\mathcal{A}$ acting on the cyclic and separating vector $\vac$ by
\begin{equation}
\label{tomita}
    Sa\vac=a^\dagger\vac.
\end{equation}
This is an unbounded antilinear operator satisfying $S^2=\mathds{1}$, and thus invertible with $S^{-1}=S$. It is also closeable so it admits a unique polar decomposition,
\begin{equation}
\label{polar_decomposition}
    S=J\Delta^{1/2},
\end{equation}
where $\Delta$ is positive and $J$ antiunitary. These are the modular operator and modular conjugation associated with $\mathcal{A}$ and $\vac$, respectively. The main result of Tomita-Takesaki theory is stated in terms of these two 
operators:
\begin{equation}
    \Delta^{is}\mathcal{A}\Delta^{-is}=\mathcal{A}\quad(s\in\mathbb{R})\quad\text{and}\quad J\mathcal{A}J=\mathcal{A}',
\end{equation}
which means that the modular operator defines a one-parameter group of automorphisms on the algebra called the \textit{modular flow}, and that the modular conjugation gives an isomorphism between the algebra and its commutant $\mathcal{A}'$. Another important property which is fundamental for our work is that $J=J^\dagger$, and thus $J^2=\mathds{1}$.

For finite-dimensional algebras acting on some factor of a product Hilbert space, the modular operator is closely related to the reduced density matrices associated with $\vac$, and the modular flow is given by the dynamics generated by these density matrices. Furthermore, the modular flow coincides with the time evolution in the case of a thermal state of unit temperature. This last fact is generalized to infinite-dimensional algebras via the so-called KMS condition and its relation to Tomita-Takesaki theory. Consider a state $\phi$ and let $\alpha:\mathbb{R}\times\mathcal{A}\to\mathcal{A}$ be a one-parameter group of automorphisms of $\mathcal{A}$ which may be thought as some dynamics on the system. One says that $\alpha$ satisfies the KMS condition with respect to $\phi$ if for every $a,b\in\mathcal{A}$ there exists a complex function $G(z)$ analytic on the strip $\im z\in(-\beta,0)$ and continuous on its closure, such that
\begin{equation}
\label{KMS}
    G(s)=\langle a\alpha_s(b)\rangle_\phi\quad\text{and}\quad G(s-i\beta)=\langle b\alpha_{-s}(a)\rangle_\phi,
\end{equation}
for $s\in\mathbb{R}$. This condition characterizes thermal states at inverse temperature $\beta$ in the infinite-dimensional setting. It turns out that the modular flow $\alpha_s(a)=\Delta^{is}a\Delta^{-is}$ satisfies the KMS condition with respect to the cyclic and separating $|\Omega\rangle$ for $\beta=1$, a fact sometimes called the \textit{modular condition}. Actually, it can be shown that it is the only one-parameter group satisfying this condition.

In order to prove the last statement another important intermediate result is used. The latter will be extensively employed in this work and states that the map $s\mapsto\Delta^{is}a\vac$ is analytic on the interior of the strip $\im s\in(-1/2,0)$ and continuous on its boundary.

There is a natural way of associating a von Neumann algebra $\mathcal{A(U)}$ to every spacetime region $\mathcal U$ in any QFT, namely taking the double commutant of the algebra of bounded operators localized in $\mathcal U$. A famous result in QFT known as the Reeh-Schlieder theorem establishes that the vacuum state $\vac$ is cyclic for any algebra $\mathcal{A}(\mathcal{U})$ such that $\mathcal{U}$ is non-empty. If the causal compliment $\mathcal{U}'$, i.e. the largest open region spacelike separated from $\mathcal{U}$, is also non-empty, then $\vac$ is also cyclic for $\mathcal{A}({\mathcal{U}'})$. Due to the fact that any cyclic vector for some algebra is separating for its commutant, one has that $\vac$ is separating for $\mathcal{A}'({\mathcal{U}'})$. Furthermore, locality imposes that $\mathcal{A}(\mathcal{U})\subseteq\mathcal{A}'({\mathcal{U}'})$ and then the vacuum vector is cyclic and separating for any algebra $\mathcal{A}(\mathcal{U})$ provided that both $\mathcal{U}$ and $\mathcal{U}'$ are non-empty. This makes possible the application of Tomita-Takesaki theory to QFT for vacuum states; but also, as we will discuss in the next section, Tomita-Takesaki theory can be applied to thermal states.

\section{The chiral fermion on the torus}\label{section3}

We consider a free massless Dirac field in $1+1$ spacetime dimensions, given by a two-component spinor $\Psi=(\Psi_+,\Psi_-)$, where $\Psi_\pm$ are the chiralities of the field. It turns out that each chirality depends on a certain combination of the spacetime coordinates, namely the null coordinates $x^\pm=t\pm x$ that give the direction of the light rays. Thus the $\Psi_\pm$ are one-variable functions $\psi_\pm(x^\pm)$, which are subject to the canonical anticommutation relations:
\begin{equation}
\label{CAR}
    \{\psi_\pm(x),\psi_\pm^\dagger(y)\}=\delta(x-y),
\end{equation}
the remaining vanishing. Strictly speaking, the objects $\psi_\pm$ are distributions and have to be smeared with a test function $f$ to give a well-defined operator, $\psi_\pm(f)=\int\dd x\ \psi_\pm(x)f(x)$. Using \eqref{CAR} one sees that the $\psi_\pm(f)$ are bounded operators and then can be used to generate the local algebras associated with an open spacetime region $\mathcal{U}$ as
\begin{equation}
    \mathcal{A(U)}=\{\psi_+(f_+),\psi^\dagger_+(f_+),\psi_-(f_-),\psi^\dagger_-(f_-)\ :\ \mathrm{supp}(f_\pm)\subseteq\pi^\pm(\mathcal{U})\}'',
\end{equation}
where $\pi^\pm(\mathcal{U})$ is the projection of the region $\mathcal{U}$ on the $x^\pm$ axis.

As in any fermionic theory, in this model the field operators localized in two spacelike separated regions $\mathcal{U}$ and $\mathcal{V}$ anticommute. But this does not mean that the associated algebras anticommute because a product of an even number of fields gives a bosonic operator, which commutes at spacelike separation with another of the same kind. So, in order to establish a relation between $\mathcal{A(U)}$ and $\mathcal{A(V)}$ we need to introduce the twist operator
\begin{equation}
    Z=\frac{\mathds{1}+iU_\pi}{1+i},
\end{equation}
where $U_\pi$ is the unitary operator associated with a U$(1)$ transformation of phase $\pi$, i.e. $U_\pi\psi_\pm U_\pi^\dagger=-\psi_\pm$. Clearly, this operator anticommutes with $\psi_\pm$ and $\psi_\pm^\dagger$, which means that for every $a\in\mathcal{A(U)}$, the product $aU_\pi$ commutes with everything in $\mathcal{A(V)}$. Then, noting that
\begin{equation}
\label{Z}
    Z\psi_\pm Z^\dagger=-i\psi_\pm U_\pi,
\end{equation}
one has that for every $a\in\mathcal{A(U)}$ the operator $ZaZ^\dagger$ commutes with everything in $\mathcal{A(V)}$. This can be promoted to a relation between the algebras
\begin{equation}
    Z\mathcal{A(U)}Z^\dagger\subseteq\mathcal{A}'(\mathcal{V}),
\end{equation}
which is the relation we were looking for. 

In this article we will primarily focus on the case of a fermion on a finite volume at finite temperature $\beta^{-1}$, i.e. on an euclidean torus of radii $L$ and $\beta$. Depending on the spin structure chosen, the field can have periodic (Ramond sector) or antiperiodic (Neveu-Schwarz sector) boundary conditions on $L$, whereas the fermi statistics impose antiperiodicity for the boundary conditions on $i\beta$. Setting $\nu=0$ for the periodic sector and $\nu=1$ for the antiperiodic one, the two-point function on this setup reads
\begin{equation}
    \label{2point}
    \langle\psi_\pm^\dagger(x)\psi_\pm(y)\rangle = \frac{1}{2\pi i} \, \frac{\sigma(x-y+L/2+i\nu\beta/2)}{\sigma(x-y-i\varepsilon)\sigma(L/2+i\nu\beta/2)} \, e^{-[\zeta(L/2)+\nu\zeta(i\beta/2)](x-y)},
\end{equation}
given in terms of the Weierstrass functions $\zeta$ and $\sigma$ (see \cite{pastras17} for an extensive review of the Weierstrass and elliptic functions) and where $\varepsilon>0$ is sent to zero after smearing. From \eqref{2point} one can calculate the two-point function for a thermal state on the line on the limit $L\to\infty$, or for the vacuum state on the circle on the limit $\beta\to\infty$, or even for the vacuum on the line taking both limits simultaneously. Also note that \eqref{2point} satisfies the stated quasiperiodicity conditions for both arguments. Furthermore, since time evolution is simply a translation of the null coordinates, the antiperiodicity property in the imaginary direction is a particular example of the KMS condition \eqref{KMS} at inverse temperature $\beta$ for $a=\psi^\dagger(x)$ and $b=\psi(y)$, as should happen for any thermal state.

We can express the two-point function \eqref{2point} as well as any other expectation value for this state using a cyclic vector in some Hilbert space via the so-called GNS construction. Roughly speaking, given any state $\phi$ acting on some von Neumann algebra $\mathcal{A}$, this construction generates a Hilbert space $\mathcal{H}$ and a representation of the algebra elements $\rho:\mathcal{A}\to\mathcal{B(H)}$ with a cyclic vector $|\Phi\rangle$ such that $\langle a\rangle_\phi=\langle\Phi|\rho(a)|\Phi\rangle$ for every $a\in\mathcal{A}$. The set $\rho(\mathcal{A})$ is itself a von Neumann algebra, and in the case where $\phi$ is a thermal state at inverse temperature $\beta$ we will denote it by $\mathcal{A}_\beta$. Remarkably, for the QFT in question it turns out that if $\phi$ is a thermal state for the global algebra $\mathcal{A}^g=\{\,\bigcup_{\mathcal{U}}\mathcal{A}({\mathcal{U}})\}''$ then the cyclic vector given by the GNS construction, which we will call $\tvac$, is cyclic and also separating for the GNS-representation of any local algebra $\mathcal{A}({\mathcal{U}})$ \cite{longo12}. This fact allows us to apply Tomita-Takesaki theory in this context. 

By writing \eqref{2point} in terms of $\tvac$, one sees that the vector valued functions $\psi_\pm(y)\tvac$ and $\psi^\dagger_\pm(y)\tvac$ are analytic on the strip $\im y\in(0,\beta)$, because its right hand side can be extended continuously in $y$ to an analytic function in this strip. This remains valid for the limits $L\to\infty$ and $\beta\to\infty$, since in any case \eqref{2point} has a pole when $y=x-i\varepsilon$.

In the finite-dimensional case, the GNS construction of a thermal state consists of a simple purification in which one doubles the Hilbert space of the theory and thus enlarges the global algebra, giving rise to a non-trivial commutant. In the infinite-dimensional setting this is also the case: alongside with $\mathcal{A}_\beta$ one can always construct a second non-trivial representation $\tilde{\mathcal{A}}_\beta$ such that $Z\tilde{\mathcal{A}}^g_\beta Z^\dagger=\big(\mathcal{A}^g_\beta\big)'$ \cite{haag67}. Furthermore, since $\mathcal{A}_\beta\mathcal{(U)}\subseteq\mathcal{A}^g_\beta$, one has
\begin{equation}
\label{second_world}Z\tilde{A}_\beta^gZ^\dagger\subseteq\mathcal{A}'_\beta(\mathcal{U}),
\end{equation}
for every $\mathcal{U}$; meaning that there is a (twisted) copy of the global algebra inside the commutant of the algebra of any open region for a thermal state. The representation $\tilde{\mathcal{A}}_\beta$ acts on the same Hilbert space as $\mathcal{A}_\beta$ provided that it is generated by operators time-evolved by a complex parameter $i\beta/2$, so \eqref{second_world} in fact also means that
\begin{equation}
\label{complex_cr}
    \{\psi_\pm(x),\psi^\dagger_\pm(y+i\beta/2)\}=0,
\end{equation}
for every $x\in\pi^\pm(\mathcal{U})$ and $y\in\mathbb{R}$.\footnote{This statement should be taken with some care: as we have remarked earlier, technically the fields $\psi_\pm(x)$ are not elements of $\mathcal{A}_\beta(\mathcal{U})$ until smeared with a test function; moreover, the field $\psi_\pm(x+i\beta/2)$ is not a bounded operator even when smeared with just any test function. The expression \eqref{complex_cr} can be made rigorous by smearing the fields with test functions with compact spectral support (see \cite{haag67}), but we will omit this detail in order to gain clarity.}

\section{Modular conjugation for the chiral fermion}\label{section4}

\subsection{Plane}

Let us start off by reviewing the method employed in \cite{nuestro_paper} to derive the modular conjugation for the vacuum state reduced to a generic multicomponent region $\mathcal U$. The null projection of $\mathcal U$ will in general be a collection of $n$ intervals on the line:
\begin{equation}
    \pi^\pm(\mathcal{U})=\bigcup_{i=1}^{n_\pm}(a_i^\pm,b_i^\pm)=V_\pm.
\end{equation}
From now on we suppress the $\pm$ indices for notational simplicity; all the results hold for both chiralities. The modular flow is non-local for $n>1$, meaning it mixes operators on $n$ points on $V$, explicitly given by \cite{reyes20}
\begin{equation}
    \label{MF_plane}
    \alpha_s(\psi(x))=2\sinh(\pi s)\sum_{i=1}^n\frac{1}{z'(x_i(s))}\frac{1}{x-x_i(s)}\psi(x_i(s)).
\end{equation}
For $u \in \mathbb{C}-V$ we conveniently define
\begin{equation}
    \label{w_plano}
    z(u) = i\pi - \int_{V} \mathrm{d}t \; \frac{1}{t-u}\,.
\end{equation}
Then for $x\in\mathbb{R}$ we take $z(x):=\lim_{\varepsilon\to0^+} z(x+i\varepsilon)=\log\of{-\prod_{i=1}^n\frac{x-a_i}{x-b_i}}$ and the $x_i(s)$ are the $n$ solutions to
\begin{equation}
    z(x_i(s)) = z(x) - 2\pi s.
\end{equation}
Clearly, \eqref{w_plano} gives an analytic continuation of $z(x)$ in $\mathbb{C}-V$. It also satisfies $0 \leq \mathrm{Im}\,z(u) \leq \pi$. This is not hard to see with the following geometrical argument: integrating in polar coordinates centered at $t=u$, when $\mathrm{Im}\,u>0$ the contour goes counterclockwise around $u$, so the imaginary part of the integral is a sum of positive angles which gives less than $\pi$. Also clearly $\mathrm{Im}\,z(x+i\varepsilon)=0$ for $x\in V$ and $\mathrm{Im}\,z(x+i\varepsilon)=\pi$ for $x\in \bar{V}$ (the complement of $V$), so the claimed property of $z(u)$ follows. This fact allows us to write the action of the modular flow on the the cyclic and separating $\vac$ for a complex parameter $s$ in the strip $\im s\in(-1/2,0)$ as a contour integral over the whole upper-half complex plane
\begin{equation}
\label{MF_plane_integral}
\alpha_s(\psi(x))\vac=\Delta^{is}\psi(x)\vac=-\frac{1}{2\pi i}\oint\dd u\,\frac{\sinh\off{\frac{\omega(x)-\omega(u)}{2}}}{\sinh\off{\frac{\omega(x)-\omega(u)}{2}-\pi s}}\frac{1}{x-u}\psi(u)\vac.
\end{equation}
The integrand is a meromorphic function of $u$ with poles given where the argument of the hyperbolic sine in the denominator vanishes. For $s$ in the strip these poles are located in the upper-half plane -- they are the modular flow trajectories $x_i(s)$ -- and applying the residue formula one gets \eqref{MF_plane} for a complex argument on the strip acting on the vacuum. This means that one can analytically continue this expression for $\im s\in(-1/2,0)$ and continuously to the strip's boundary. 

Since \eqref{MF_plane} is also valid using $\psi^\dagger$ instead of $\psi$, taking $s=-i/2$ we can obtain the action of $\Delta^{1/2}$ over the vector $\psi^\dagger(x)\vac$ and then relate it to the modular conjugation via the Tomita operator $S$. Indeed, using \eqref{tomita}, \eqref{polar_decomposition} and the fact that $J^2=\mathds{1}$ one has
\begin{equation}
\label{Delta_to_J}    \Delta^{1/2}\psi^\dagger(x)\vac=J\psi(x)\vac.
\end{equation}
One can promote this vectorial expression to an operatorial identity as we describe in the following. Evaluating $s=-i/2$ takes the trajectories $x_i(s)$ to the $n$ points $\bar{x}_i$ that are solutions to
\begin{equation}
\label{x_bar_plane}
    z(\bar{x}_i)=z(x)+i\pi.
\end{equation}
From our previous analysis of $\mathrm{Im}\,z(u)$, note that the $\bar{x}_i$ are located in $\bar{V}$, and thus the fields $\psi^\dagger(\bar{x}_i)$ that appear acting on the vacuum in $\Delta^{1/2}\psi^\dagger(x)\vac$ anticommute with everything in $\mathcal{A(U)}$. Using \eqref{Z}, and the fact that $U_\pi\vac=\vac$, we can replace each of these fields acting on the vacuum with $iZ\psi^\dagger(\bar{x}_i)Z^\dagger\vac$, such that the left hand side of \eqref{Delta_to_J} becomes a linear combination of elements of $\mathcal{A}'(\mathcal{U})$ acting on the vacuum. The right hand side can also be rewritten in terms of an element of $\mathcal{A}'(\mathcal{U})$ acting on the vacuum as $J\psi(x)J\vac$, and then the separantability of $\vac$ implies a relation between the operators on both sides of the equation. This relation gives the explicit action of the modular conjugation:
\begin{equation}
\label{J_plano}
    J\psi(x)J=2\sum_{i=1}^n\frac{1}{z'(\bar{x}_i)}\frac{1}{x-\bar{x}_i}Z\psi^\dagger(\bar{x}_i)Z^\dagger.
\end{equation}

This operator inherits the same non-local behaviour of the modular flow: for $n>1$ it consists of a combination of fields defined on distinct points on $\bar{V}$, whereas for $n=1$ it is given by a completely local transformation.
It is also discontinuous when $z(x)=0$, which happens once within each interval. So when $x$ approaches the roots of $z$ from left or right, there is one solution of \eqref{x_bar_plane} that discontinuously jumps from $+\infty$ to $-\infty$.

\subsection{Torus}

The same procedure discussed in the previous paragraphs can be employed to derive the action of the modular conjugation whenever the modular flow is known. For the chiral fermion on multicomponent regions on the torus, the modular flow has been recently computed in \cite{reyes20} giving
\begin{equation}
\label{MF_torus}
    \alpha_s(\psi(x))=\frac{2\pi}{\beta} \sinh(\pi s)\sum_{k\in\mathbb{Z}}\sum_{i=1}^n\frac{1}{z'(x_{ik}(s))}\frac{(-1)^{k\nu}}{\sinh\left[\frac{\pi}{\beta}(x-x_{ik}(s)+kL)\right]} \psi(x_{ik}(s)),
\end{equation}
where we will define the function $z$ as
\begin{equation}
    \label{z_torus}
    z(u) = i\pi \; + \; \int_V \mathrm{d}t \left[ \zeta(u-t) - u \frac{\zeta(i\beta/2)}{i\beta/2} \right] \,,
\end{equation}
and calling $\ell = \int_V \dd t$, the total length of $V$, the $x_{ik}(s)$ that appear in \eqref{MF_torus} are the modular flow trajectories, which are solutions to the equation
\begin{equation}
\label{MF_torus_tray}
    z(x_{ik}(s))=z(x)+\frac{2\pi\ell }{\beta}k-2\pi s\qquad(k\in\mathbb{Z}).
\end{equation}
Within each interval $(a_i,b_i)$, $z$ is a monotonically increasing function ranging from $-\infty$ to $+\infty$ \cite{DyG_19}. Hence for each $k$ there are $n$ solutions $x_{ik}(s)$.

Note that the flow \eqref{MF_torus} is non-local as well as the one presented in the previous subsection \eqref{MF_plane}. But a prominent difference appears among them when $n=1$: in this case the latter becomes completely local, whereas the sum over $k\in\mathbb{Z}$ in the former makes it a non-local operator even for a single interval. We expect this fact to be present also in the modular conjugation for the torus.
 
\begin{figure}[ht!]
    \centering
    \captionsetup{justification=centering}
    \includegraphics[width=0.55\textwidth]{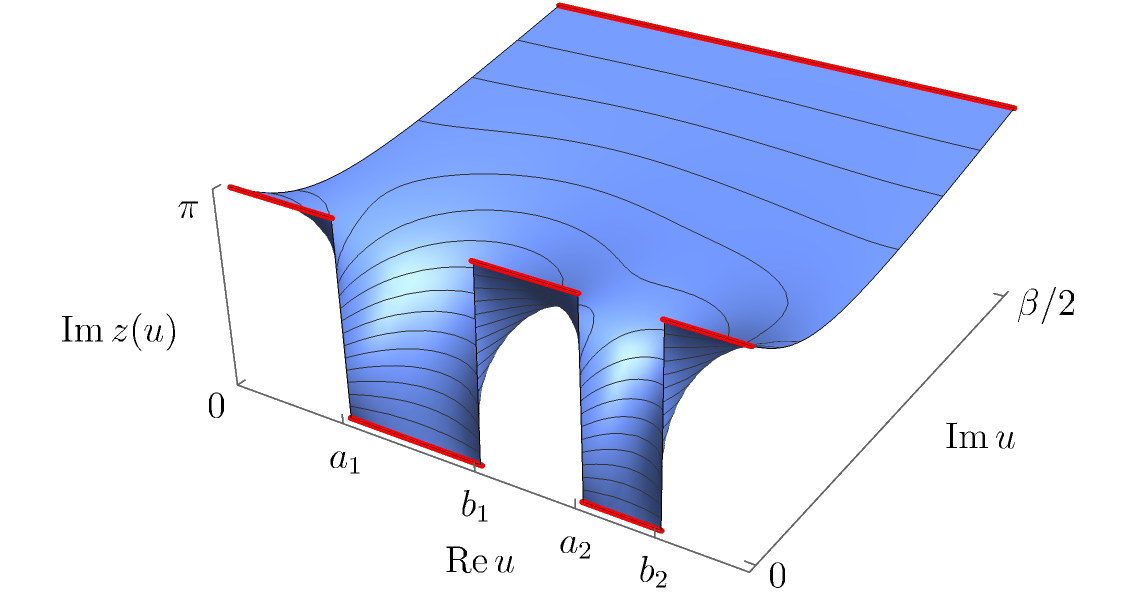}
    \caption{$\mathrm{Im}\,z(u)$ for $u\in(-L/2,L/2) \times i(0,\beta/2)$ in the case of $n=2$ intervals. We highlighted in red the regions where $\mathrm{Im}\,z=0$ or $\im z=\pi$; these are respectively $V$ or $\bar{V}$ and $\mathrm{Im}\,u=\beta/2$.}
    \label{fig:plotimz}
\end{figure}

\begin{figure}[ht!]
    \centering
    \captionsetup{justification=centering}
    \includegraphics[width=0.9\textwidth]{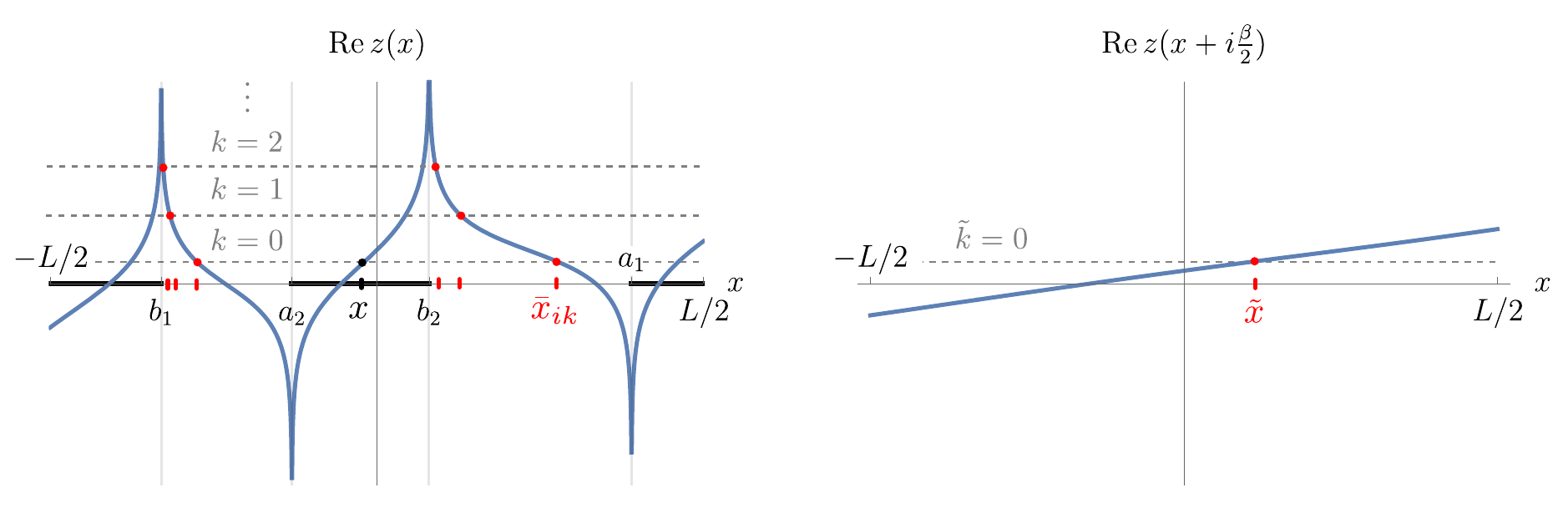}
    \caption{$\mathrm{Re}\,z(u)$ for $\mathrm{Im}\, u = 0$ (left) and $\mathrm{Im}\, u = \beta/2$ (right) in the case of $n=2$ intervals. In the left panel we also show some solutions to \eqref{eq:polos} for $x\in V$ in red, whose positions are given by the intersections of the real part of $z$ with horizontal lines of height $z(x)+2\pi\ell k/\beta$ that occur in $\bar{V}$. In the right panel, note that the real part of $z(x+i\beta/2)$ is monotonically increasing and thus there exists a complex solution to \eqref{x_k0} with $\tilde{k}=0$.}
    \label{fig:plotrez}
\end{figure}

Mimiking the procedure employed for the plane, we are interested in studying the analytic continuation of \eqref{MF_torus} acting on the cyclic and separating $\tvac$ when $-1/2<\mathrm{Im}\,s<0$.  In order to do so, we begin by analyzing some properties of $z(u)$. First, it is a quasiperiodic function $z(u+L)=z(u)+\frac{2\pi\ell}{\beta}$. Secondly, a bound for its imaginary part analogous to that of the plane holds, which is key for our analytic continuation. In figure \ref{fig:plotimz} we show that in the interior of the region $0 < \mathrm{Im}\,u < \beta/2$ we have $0 < \mathrm{Im}\,z(u) < \pi$. On its boundary we have that $\mathrm{Im}\,z(u) = 0$ for $u\in V$, while $\mathrm{Im}\,z(u) = \pi$ for $u\in\bar{V}$ and also when $\mathrm{Im}\,u = \beta/2$. We prove these statements in appendix \ref{sec:appendixzeta}. Finally, $\mathrm{Re}\,z$ is a monotonically decreasing function within each interval in $\bar{V}$, ranging from $+\infty$ to $-\infty$ and monotonically increasing from $-\infty$ to $+\infty$ in the infinite line $\mathrm{Im}\,u = \beta/2$ (note that it is continuous). We plot the real part in figure \ref{fig:plotrez}.

Now, consider the vector-valued function
\begin{equation}
\label{MF_torus_integral}
|\varphi_x(s)\rangle=\frac{i}{2\beta} \oint_{\partial\Sigma} \dd u\; \frac{\sinh\left[\frac{z(x)-z(u)}{2}\right]}{{\sinh\left[\frac{z(x)-z(u)}
    {2}-\pi s\right]}} \frac{1}{\sinh\left[\frac{\pi}{\beta}(x-u)\right]} \psi(u)\tvac,
\end{equation}
where the contour $\partial\Sigma$ is the boundary of the infinite region $\Sigma=\{u\in\mathbb{C}\ :\ 0<\im u<\beta/2\}$. Note that the integrand is a meromorphic function of $u$, and then the integral can be computed by residues. Since $\im z\in(0,\pi)$ for $u\in\Sigma$, taking $s$ in the strip $\im s\in(-1/2,0)$, the poles of the integrand are located where the argument of the first hyperbolic sine in the denominator vanishes. This is when $z(u)=z(x)+2\pi s$ and gives solutions $u_{ik}(s)$ on the interior of the whole region $\Sigma$. Then we make the identification $u_{ik}(s) + kL \equiv x_{ik}(s)$ for $k\in\mathbb{Z}$, such that all the $x_{ik}(s)$ lie on the same fundamental domain of the torus. Taking into account the quasiperiodicity of $\psi$ and $z$, one sees that under this identification the set of poles precisely match the modular trajectories \eqref{MF_torus_tray}. Then applying the residue formula one arrives to
\begin{equation}
    |\varphi_x(s)\rangle=\frac{2\pi}{\beta} \sinh(\pi s)\sum_{k\in\mathbb{Z}}\sum_{i=1}^n\frac{1}{z'(x_{ik}(s))}\frac{(-1)^{k\nu}}{\sinh\left[\frac{\pi}{\beta}(x-x_{ik}(s)+kL)\right]} \psi(x_{ik}(s))\tvac.
\end{equation}
Some comments are in order. In the first place, in the strip $-1/2<\mathrm{Im}\,s<0$ there are no poles on the integration contour, hence the function $|\varphi_x(s)\rangle$ is analytic in this strip and continuous on its boundary. For $\mathrm{Im}\,s\to0$ we recover \eqref{MF_torus} acting on $\tvac$ as expected. Similarly, we can take the limit $s\to-i/2$ in order to obtain the action of $\Delta^{1/2}$. In that case, the imaginary part of $z$ determines that the solutions of the complex modular flow trajectories are located either in $\bar{V}$ or in $\mathrm{Im}\,u=\beta/2$. To make this distinction explicit, we shall rewrite them as $\bar{x}_{ik}$ or $\tilde{x} + i\tfrac{\beta}{2}$, where
\begin{align}
    z(\bar{x}_{ik}) & = z(x)+\frac{2\pi\ell}{\beta}k+i\pi \label{eq:polos}\,,\\
    z(\tilde{x}+i\tfrac{\beta}{2}) & = z(x)+\frac{2\pi\ell}{\beta}\tilde{k}+i\pi \label{x_k0}\,,
\end{align}
with $\bar{x}_{ik}\in\bar{V}$ and $\tilde{x}\in\mathbb{R}$ within a fundamental period of the torus and some given $\tilde{k}$. Let us describe the solutions to these equations. Because of the monotonicity of $z$, \eqref{eq:polos} has $n$ solutions $\bar{x}_{ik}$ for each $k\in\mathbb{Z}$, given we pick convenient coordinates\footnote{Even though the set of solutions for all $k$ in any case is the same, equation \eqref{eq:polos} has the cumbersome feature that if the borders of the chosen fundamental period are in $\bar{V}$, it will have $n-1$ solutions for some particular $k$.}. Graphically, one can find these solutions by inspecting the plot of the real part of $z$ (see the left panel of figure \ref{fig:plotrez}): the several intersections of this plot with horizontal lines of height $z(x)+\frac{2\pi\ell}{\beta}k$ that occur in the complement $\bar{V}$, where $\im z=\pi$, give the positions of the $\bar{x}_{ik}$. Similarly to the modular flow, the $\bar{x}_{ik}$ accumulate near the endpoints of the intervals.

Remarkably however the $\bar{x}_{ik}$ are not the complete set of solutions. Because $\mathrm{Re}\,z(x+i\tfrac{\beta}{2})$ grows monotonically, there always exists one $k=\tilde{k}$ such that \eqref{x_k0} gives exactly one complex solution. 

Moving on, remembering \eqref{Delta_to_J} and that \eqref{MF_torus} is still valid when replacing $\psi$ by $\psi^\dagger$, we can compute the action of the modular conjugation from $\Delta^{1/2}\psi^\dagger(x)\tvac$ as we did for the plane. But taking into account the previous discussion about the complex solution we have to be more careful. In our derivation for the plane we used the fact that all the solutions to \eqref{x_bar_plane} were located in $\bar{V}$ and thus the corresponding operators anticommuted with $\mathcal{A(U)}$. This is true for the $\bar{x}_{ik}$ satisfying \eqref{eq:polos}, but what happens with the term in $\Delta^{1/2}\psi^\dagger(x)\tvac$ associated to the complex solution \eqref{x_k0}? It is worth remembering that the fields $\psi(\tilde{x}+i\beta/2)$ can be thought as being elements of the representation $\tilde{\mathcal{A}}_\beta^g$ introduced at the end of the previous section, so we have \eqref{second_world}, or more explicitly \eqref{complex_cr}. Then, the procedure we did for the plane follows and we finally arrive to
\begin{align}
    \label{J_toro}
    J\psi(x)J=\frac{2\pi}{\beta}\Bigg\{\sum_{k\in\mathbb{Z}}\sum_{i=1}^n \frac{1}{z'(\bar{x}_{ik})}&\frac{(-1)^{k\nu}}{\sinh\off{\frac{\pi}{\beta}\of{x-\bar{x}_{ik}+kL}}}Z\psi^\dagger(\bar{x}_{ik})Z^\dagger\,+\notag\\
    &\qquad+\frac{1}{z'(\tilde{x}+i\frac{\beta}{2})}\frac{(-1)^{\tilde{k}\nu}i}{\cosh\off{\frac{\pi}{\beta}\big(x-\tilde{x}+\tilde{k}L\big)}}Z\psi^\dagger(\tilde{x}+i\tfrac{\beta}{2})Z^\dagger\Bigg\}.
\end{align}
This is the main result of our work, with the complex modular trajectories given by \eqref{eq:polos}, \eqref{x_k0} and \eqref{z_torus}. Note that, as we anticipated, the modular conjugation shares the same non-locality as the modular flow, which is present even for $n=1$. Furthermore, it exhibits another novelty when compared to the result for the plane: the appearance of the complex solutions $\tilde{x}+i\beta/2$, which we explicitly display in \eqref{J_toro} and is related to the non-triviality of the commutant for the representation of the operator algebra in a thermal state, as we show in the next subsection. It occurs that this modular conjugation not only maps $\mathcal{A}_\beta(\mathcal{U})$ to $\mathcal{A}_\beta(\mathcal{U}')$, but it also maps it to $\tilde{\mathcal{A}}_\beta^g$. This behaviour has been discussed previously in \cite{longo12,borchers99,schroer99} or more recently in \cite{tonni22}, and some authors interpret that the representation $\tilde{A}_\beta^g$ acts as a ``second world'', analogous to the copy of the theory one introduces to perform a purification in a finite-dimensional setting.

\subsection{The second world}

There are a couple of things that do not come obvious by looking at the expression \eqref{J_toro}, especially due to the presence of the complex solution $\tilde{x}+i\beta/2$. First of all, one can ask for a more transparent interpretation of its emergence and exactly what is its relation to the ``second world''. On the other hand, it also makes it difficult to immediately guess the action of $J$ over the commutant algebra, which is pretty straightforward in the plane case. We aim to clarify both issues in this section.

In order to gain some insight on the complex solution, we will study a simple limit of our result. Let us consider a single interval such that $\ell\to L$, that is when $V$ becomes the whole circle, and thus the algebra of operators becomes the global algebra. Since the global algebra is of type I it admits a thermal density matrix $\rho_\beta$, and we can perform a purification by coupling the Hilbert space of the theory $\mathcal{H}$ to a copy $\tilde{\mathcal{H}}$ such that $\rho_\beta=\mathrm{tr}_{\tilde{\mathcal{H}}}\tvac\langle\Omega_\beta|$, where $\tvac\in\mathcal{H}\otimes\tilde{\mathcal{H}}$. Now, $\mathcal{A}_\beta^g$ and $\tilde{\mathcal{A}}_\beta^g$ are the algebras of operators solely acting on $\mathcal{H}$ or $\tilde{\mathcal{H}}$, respectively; and they are clearly each others commutants. Thus, if we represent the elements of $\mathcal{A}_\beta^g$ as generated by fields $\psi(x)$ with $x\in(-L/2,L/2)$, let us use the modular conjugation to probe how the elements of $\tilde{\mathcal{A}}_\beta^g$ look like. In this setting we have
\begin{equation}
    z(x)=\frac{2\pi}{\beta}x,
\end{equation}
up to an additive constant which is irrelevant. The modular flow then coincides with time evolution of parameter $-\beta s$ and hence the modular conjugation maps $x\mapsto x+i\beta/2$. In fact this is the only solution to \eqref{MF_torus_tray} in this limit. Replacing this in \eqref{J_toro} one arrives to
\begin{equation}
\label{J_global}    J\psi(x)J=iZ\psi^\dagger(x+i\beta/2)Z^\dagger.
\end{equation}
So we learn that the ``second world'' algebra is naturally given by fields on points with imaginary part equal to $\beta/2$. This is what the complex solution on \eqref{J_toro} represents. Note that here $\tilde{x}$ coincides with $x$ since we were working without a partition of the circle, but the effect of taking $\bar{V}$ non-empty is that $\tilde{x}$ in general does not coincide with $x$.

\begin{figure}[h]
    \centering
    \includegraphics[width=0.6\textwidth]{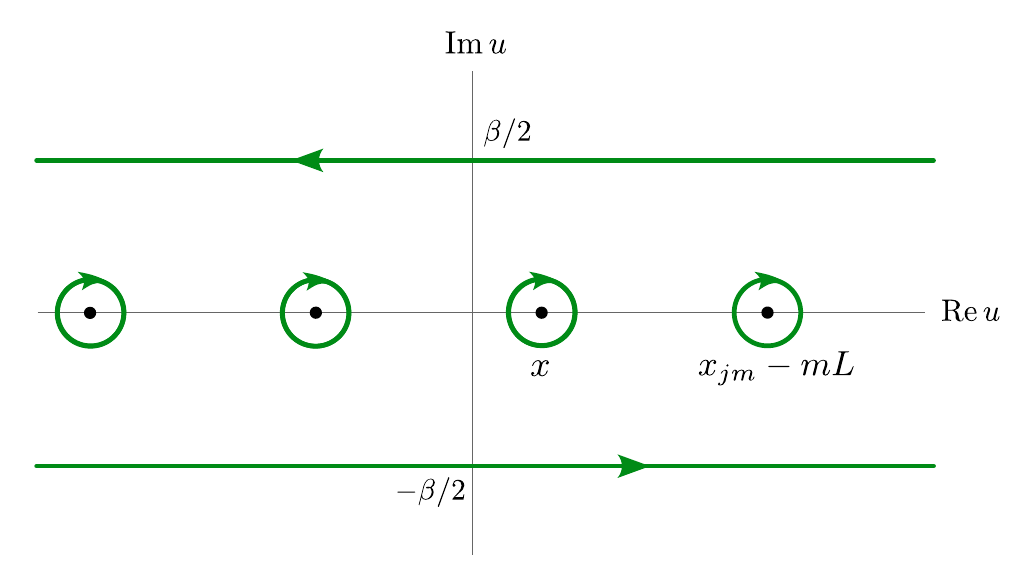}
    \caption{Integration contour $\Gamma$ of eq. \eqref{J^2_int}.}
    \label{fig:gamma}
\end{figure}

Let us now address the action of the modular conjugation over $\mathcal{A}'_\beta(\mathcal{U})$. The modular conjugation of commuting algebras should coincide \cite{witten18}, but note that \eqref{J_toro} only tells us how to act over operators in $\mathcal{A}_\beta(\mathcal{U})$, so the action over the commutant could be different and we would like to find it out. In order to do so, we will use the fact that this is an involutive operator, i.e. $J^2=\mathds{1}$, as well as other properties. First of all, note that \eqref{J_toro} implies that $\mathcal{A}_\beta'(\mathcal{U})\subseteq Z\big(\mathcal{A}_\beta(\mathcal{U}')\lor\tilde{\mathcal{A}}_\beta^g\big)Z^\dagger$ and since the opposite inclusion also holds, it can actually be promoted to an equality.\footnote{The fact that ${\mathcal{A}'}_{\!\!\beta}(\mathcal{U})\neq Z\mathcal{A}_\beta(\mathcal{U}')Z^\dagger$ may look like a violation of the so-called Haag duality. But one has to remember that the $\mathcal{A}_\beta$'s are the representations of the algebra induced by the $\beta$-thermal state, whereas Haag duality is usually evaluated over the vacuum representation, which gives the natural Hilbert space of the theory in question.} Whenever two algebras are unitarily equivalent then the corresponding modular conjugations are unitarily related, so we have $J=Z\hat{J}Z^\dagger$ \cite{nuestro_paper, foit83}, where $\hat{J}$ is  associated with $\mathcal{A}_\beta(\mathcal{U}')\lor\tilde{\mathcal{A}}_\beta^g$. Thus, knowing the action of $\hat{J}$ over $\psi(\bar{x})$ for $\bar{x}\in\bar{V}$ or $\im{\bar{x}}=\beta/2$ yields the result we are after. As an educated guess, we propose an action similar to \eqref{J_toro}:
\begin{equation}
\label{J_hat}
    \hat{J}\psi(\bar{x})\hat{J}=\frac{2\pi}{\beta}\sum_{jm}\frac{1}{\hat{z}'(x_{jm})}\frac{(-1)^{k\nu}}{\sinh\off{\frac{\pi}{\beta}\of{\bar{x}-x_{jm}+mL}}}Z\psi^\dagger(x_{jm})Z^\dagger,
\end{equation}
such that the $x_{jm}$ are solutions to
\begin{equation}
    \hat{z}(x_{jm})=\hat{z}(\bar{x})-\frac{2\pi\ell}{\beta}m+i\pi,
\end{equation}
for a suitable $\hat{z}$ defined as $\hat{z}=-z+i\pi$. This is a reasonable ansatz because if one takes $\bar{x}$ as a solution to \eqref{eq:polos} or \eqref{x_k0} for some $k\in\mathbb{Z}$ then one has
\begin{equation}
    z(x_{jm})=z(x)+\frac{2\pi\ell}{\beta}(k+m),
\end{equation}
which means that $x_{jm}\in V$. In particular it means that $x_{jm}=x$ for some $j$ and $m$. Then, at least geometrically, we see that the modular conjugation results involutive. Let us check that this choice also means $J^2=\mathds{1}$ as an operator. Using \eqref{J_toro} and \eqref{J_hat} we have
\begin{equation}
    J^2\psi(x)J^2=\frac{4\pi^2}{\beta^2}\sum_{ikjm}\frac{1}{z'(\bar{x}_{ik})}\frac{1}{z'(x_{jm})}\frac{(-1)^{k\nu}}{\sinh\off{\frac{\pi}{\beta}\of{x-\bar{x}_{ik}+kL}}}\frac{(-1)^{m\nu}}{\sinh\off{\frac{\pi}{\beta}\of{\bar{x}_{ik}-x_{jm}+mL}}}\psi(x_{jm}),
\end{equation}
where we have used that $Z^2=U_\pi$ and that $\hat{z}'=-z'$. Also this time we are including the complex solution $\tilde{x} + i\beta/2 $ in the sum over $i$ and $k$; our argument will be independent of this distinction. Using the residue formula, this expression admits an integral representation along the same curve as \eqref{MF_torus_integral}, but considering the analiticity of the integrand the region can be extended, yielding
\begin{align}
\label{J^2_int}
    J^2\psi(x)J^2&=-\frac{\pi}{   
    \beta^2}\sum_{jm}\frac{(-1)^{m\nu}}{z'(x_{jm})}\psi(x_{jm})\ \times\notag\\
    &\qquad\qquad\times\ \oint_\Gamma\dd u\,\frac{\sinh\off{\frac{z(x)-z(u)}{2}}}{\sinh\off{\frac{z(x)-z(u)}{2}+\frac{i\pi}{2}}}\frac{1}{\sinh\of{\frac{\pi}{\beta}(x-u)}}\frac{1}{\sinh\of{\frac{\pi}{\beta}(u-x_{jm}+mL)}},
\end{align}
for the curve $\Gamma$ shown in figure \ref{fig:gamma}. The integral over the horizontal lines vanishes due to the $i\beta-$periodicity of the integrand, so we are only left with the integral along the loops, which is determined by the poles at $x$ and $x_{jm}-mL$. It turns out that all the residues vanish except for the one corresponding to $m=0$ and $j$ such that $x_{j0}=x$. So we can drop the sum and study the non-vanishing term as a limit, which gives
\begin{equation}
    J^2\psi(x)J^2=\psi(x).
\end{equation}
This ultimately justifies our guess \eqref{J_hat}, which in turn gives the correct action of $J$ over $\mathcal{A}'_\beta(\mathcal{U})$.

\subsection{Limits}

\begin{figure}[ht!]
    \centering
    \captionsetup{justification=centering}
    \includegraphics[width=0.9\textwidth]{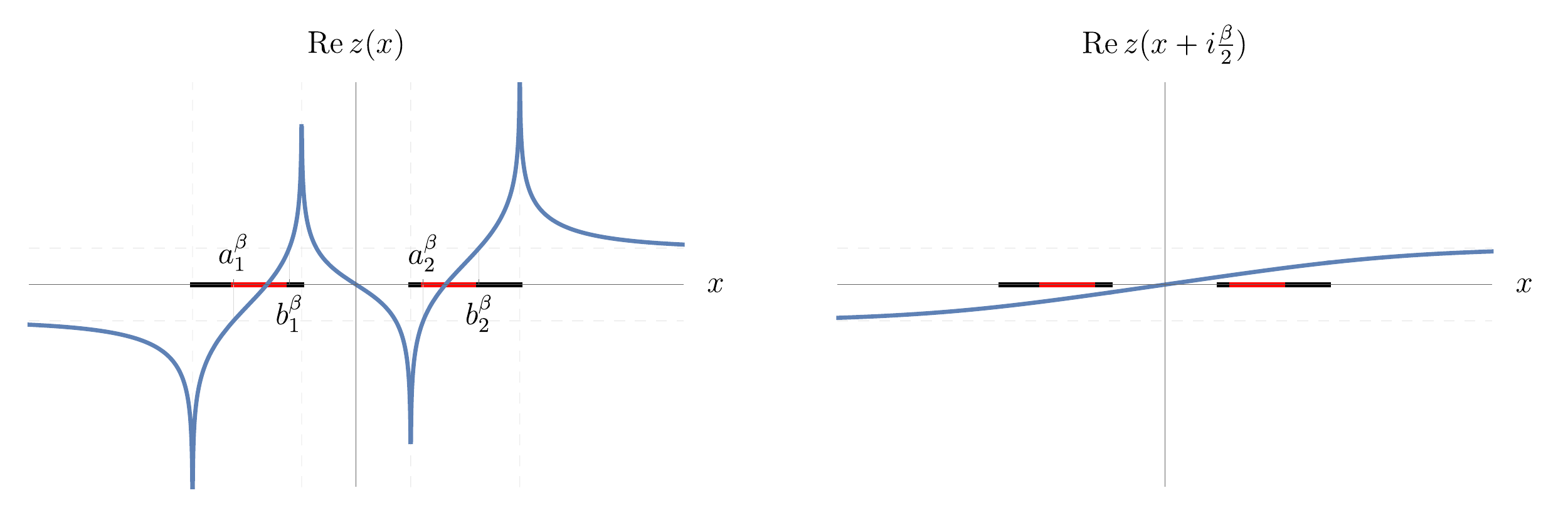}
    \caption{$\mathrm{Re}\,z(x)$ (left) and $\mathrm{Re}\,z(x+i\beta/2)$ (right) in the limit $L\to\infty$ for the case $n=2$. The thick black lines represent the two intervals of $V$, and the smaller intervals in red are $(a_1^{\beta},b_1^{\beta}) \cup (a_2^{\beta},b_2^{\beta}) = V_{\beta}$. For $x \in V_{\beta}$ we see that there are $n-1$ real solutions and one complex solution.}
    \label{fig:zthermal}
\end{figure}

Let us now study some relevant limits of our main result. We will begin by taking the limit $L\to\infty$ while keeping $\beta$ finite. In this case, after integration \eqref{z_torus} becomes
\begin{equation}
\label{z_cylinder_th}
    z(u)=\log\off{-\prod_{i=1}^n\frac{\sinh\big(\frac{\pi}{\beta}(u-a_i)\big)}{\sinh\big(\frac{\pi}{\beta}(u-b_i)\big)}},
\end{equation}
where we have dropped an irrelevant real constant and a term linear in $u/L$. Also, since every term with $k\neq0$ in \eqref{J_toro} vanishes in the limit, we only care about the $n$ solutions $\bar{x}_i=\bar{x}_{i0}$ to \eqref{eq:polos}. As we can see by inspecting the real part of $z$ (see figure \ref{fig:zthermal}), this time the complex solutions do not exist for every $x$, but only for a subset $x\in V_\beta \subseteq V$, given by a collection of smaller intervals $(a_i^{\beta},b_i^{\beta}) \subseteq (a_i,b_i)$. In that case we have $n-1$ solutions $\bar{x}_i$ in $\bar{V}$, one for each interval $(b_i,a_{i+1})$ and the complex solution $\tilde{x}+i\beta/2$. On the other hand when $x \in V-V_{\beta}$, we have $n$ solutions $\bar{x}_i$ in $\bar{V}$ as we had in the plane, with $\bar{x}_n$ either in $(-\infty,a_1)$ or $(b_n,\infty)$. We find
\begin{align}
\label{J_cylinder_thermal}
    J\psi(x)J = \frac{2\pi}{\beta} \Bigg\{ \sum_{i=1}^{n-1} \frac{1}{z'(\bar{x}_i)} & \frac{1}{\sinh\off{\frac{\pi}{\beta}(x-\bar{x}_i)}} Z\psi^\dagger(\bar{x}_i)Z^\dagger \, + \notag\\
    & \qquad + \chi_{\bar{\beta}}(x) \frac{1}{z'(\bar{x}_n)} \frac{1}{\sinh\off{\frac{\pi}{\beta}(x-\bar{x}_n)}} Z\psi^\dagger(\bar{x}_n)Z^\dagger \, + \notag\\
    & \qquad + \chi_{\beta}(x) \frac{1}{z'(\tilde{x}+i\frac{\beta}{2})}\frac{i}{\cosh\off{\frac{\pi}{\beta}\of{x-\tilde{x}}}}Z\psi^\dagger(\tilde{x}+i\tfrac{\beta}{2})Z^\dagger\Bigg\},
\end{align}

\noindent where $\chi_{\beta}(x)$ and $\chi_{\bar{\beta}}(x)$ are the characteristic functions of $V_\beta$ and $V - V_\beta$ respectively. This result represents the modular conjugation for a thermal state of the fermion on the line reduced to $n$ intervals. For $n=1$ its geometrical action was studied in \cite{tonni22}, and in fact we have checked that the map $x\mapsto\bar{x}$ there presented satisfies $z(\bar{x})=z(x)+i\pi$, with $z$ given by \eqref{z_cylinder_th}. For a single interval, the subset $V_\beta$ simply becomes a sub-interval $(a^\beta,b^\beta)\subseteq V$, and the authors of \cite{tonni22} also explicitly found it. We can recover their result noting that \eqref{x_k0} is satisfied with $\tilde{k}=0$ only if $-\pi\ell/\beta\leq z(x)\leq\pi\ell/\beta$ and the boundaries of the region where this occurs take the values
\begin{equation}
    a^\beta=-\frac{\beta}{2\pi}\log\off{\frac{e^{-2\pi a/\beta}+e^{-2\pi b/\beta}}{2}}\qquad\text{and}\qquad b^\beta=\frac{\beta}{2\pi}\log\off{\frac{e^{2\pi a/\beta}+e^{2\pi b/\beta}}{2}},
\end{equation}
that precisely match the expressions given in \cite{tonni22}.

By taking the limit $\beta\to\infty$ over the previous results we should recover the plane case. Indeed, in this limit the function $z(u)$ given by \eqref{z_cylinder_th} simply becomes \eqref{w_plano} and the expression \eqref{J_cylinder_thermal} becomes almost identical to \eqref{J_plano}, except for the term associated to the complex solution. However, remind this term only appears for $x\in V_\beta$. Since $\mathrm{Re}\,z(x+i\beta/2) \to 0$, each sub-interval $(a_i^\beta,b_i^\beta)$ collapses to a point given by $z(x_i)=0$, which is where the map $x\mapsto\bar{x}$ for the plane is discontinuous.

Lastly, let us take the limit $\beta\to\infty$ with $L$ remaining finite. Now \eqref{z_torus} becomes after integration
\begin{equation}
        z(u)=\log\off{-\prod_{i=1}^n\frac{\sin\of{\frac{\pi}{L}(a_i-u)}}{\sin\of{\frac{\pi}{L}(b_i-u)}}} + \frac{2\pi\ell}{L} \frac{u}{\beta},
\end{equation}
again dropping an irrelevant real constant. Naturally the second term can also be dropped for any bounded $u$, but here we have non-vanishing contributions for $|u|\sim\beta$. For the terms corresponding to the real solutions \eqref{eq:polos}, one has to split the sum in $k$ into two separate contributions as explained in \cite{DyG_19}: The terms with $\ell |k| \ll \beta$ concentrate around $\bar{x}_{i} = \bar{x}_{i0}$ independently of $k$. On the other hand the terms with $\ell |k| \sim \beta$ behave drastically different depending on the boundary conditions. In the Neveu-Schwarz sector consecutive terms in $k$ do not contribute because of antiperiodicity, while in the Ramond sector the real solutions $\bar{x}_{ik}$ become densely distributed and give rise to a completely non-local contribution. Finally, the modular conjugation has the contribution of the complex solution, which in this limit has real part given by
\begin{equation}
    \tilde{x}+\tilde{k}L = \frac{L\beta}{2\pi\ell} z(x) \,.
\end{equation}
This equation is ill-defined when $\beta\to\infty$, so we carefully proceed as follows. We see that $\tilde{x}$ oscillates rapidly around the circle as we shift $x$, suggesting that one should take an average around the circle rather than a fixed $\tilde{x}$. In fact, we can consider instead an operator $\psi(f)$ where $f$ is a test function with support narrowly centered around $x$. Then, for large $\beta$, the contribution of the complex solution is
\begin{equation}
    \label{eq:bdpsmeareado}
    \frac{1}{2\cosh\left[ \frac{L}{2\ell} z(x) \right]} \int_{-\infty}^{\infty} \mathrm{d}t \; f(t) \; \int_{-L}^{L} \frac{\mathrm{d}y}{\ell} \; \psi(y+i\tfrac{\beta}{2}) |\Omega\rangle\,.
\end{equation}
We get indeed a contribution averaged over the circle on the second world. Up to a normalization factor this does not depend on $f$, which can now be taken arbitrarily narrow and ultimately justifies our previous guess. Notably, because of antiperiodicity (\ref{eq:bdpsmeareado}) vanishes in the Neveu-Schwarz sector, so there is a second world contribution only in the Ramond sector. Summing up, we have that the $\beta\to\infty$ limit of the modular conjugation for the antiperiodic sector is
\begin{equation}
    J\psi(x)J=\frac{2\pi}{L} \sum_{i=1}^{n} \frac{1}{z'(\bar{x}_i)} \csc\off{\frac{\pi}{L}\of{x-\bar{x}_i}} Z\psi^\dagger(\bar{x}_i)Z^\dagger \,,
\end{equation}
while for the periodic one
\begin{align}
    \label{eq:JRamond}
    J\psi(x)J  = \frac{2 \pi}{L} &\sum_{i=1}^n \frac{1}{z'(\bar{x}_i)} \cot\left[\frac{\pi}{L}(x-\bar{x}_i)\right] Z\psi^{\dagger}(\bar{x}_i)Z^{\dagger} \; - \notag\\
    &\qquad - \int_{\bar{V}} \frac{\mathrm{d}y}{\ell} \; \frac{1}{\sinh\left[\frac{L}{2\ell}\left( z(x)-z(y)+i\pi \right)\right]} \; Z\psi^{\dagger}(y)Z^{\dagger} \; + \notag\\
    &\qquad + \frac{i}{\cosh\left[ \frac{L}{2\ell} z(x) \right]} \int_{-\frac{L}{2}}^{\frac{L}{2}} \frac{\mathrm{d}y}{\ell} \; Z\psi^{\dagger}(y+i\beta/2)Z^{\dagger} \,.
\end{align}
In the second term of \eqref{eq:JRamond} there is an implicit principal value in the integral over $y$, due to the fact that we have summed the $\ell |k| \ll \beta$ terms separately. We also mention that we have obtained the same result by first taking the limit in \eqref{MF_torus_integral}. There we use quasiperiodicity and sum in $k$ in a similar fashion and then compute the modular conjugation. Doing so we obtain the third term of \eqref{eq:JRamond} without invoking a test function.

The absence or presence in either case of a second world term is remarkable. It is not surprising however if we remember that the zero temperature state of the Ramond sector is a mixture of two independent vacuum states due to the presence of a zero mode. We still have to somehow purify the state, and thus enlarge the algebra. To that end we can think of it as a thermal state with arbitrarily large inverse temperature $\beta$ and then use the GNS construction as before. In fact this is the only surviving term of the modular conjugation in the limit $\ell \to L$. There the algebra $\mathcal{A}(\mathcal{U})$ becomes a global algebra with trivial commutant in the Neveu-Schwarz sector, while in the Ramond case the algebra is enlarged and the modular conjugation simply maps the algebra to
\begin{equation}
    J\mathcal{A}(\mathcal{U})J = Z\tilde{\mathcal{A}}^gZ^\dagger = \mathcal{A}'(\mathcal{U})\,,
\end{equation}
as it should be. The non-pureness of the Ramond zero temperature state also manifests similarly as a $\log 2$ contribution to the entropy, as shown in \cite{DyG_19,klich17}. 

\section{Final comments}\label{section5}

In this paper we computed a new modular conjugation, namely that of a free massless fermion in $1+1$ dimensions for multicomponent regions on the torus. We followed an analogous method to the one we had employed in the plane in our previous work. As expected, this modular conjugation shares very similar geometrical features with the modular flow, the most remarkable one being the non-locality even present for the case of a single interval. We also studied explicitly the limits of this result when the region is the whole circle and when either one of the periods or both approach infinity, recovering in the latter case the previously known result.

When compared to the case of the plane however, a novel contribution to the modular conjugation associated to a complex solution to the complex modular flow trajectories appears. We argued that the operators $\psi(\tilde{x}+i\beta/2)$ should be associated to the algebra $\tilde{\mathcal{A}}_\beta^g$, sometimes identified as a ``second world", related to the GNS construction that comes into play with thermal states. The modular conjugation must map the algebra to its commutant, and since the GNS construction enlarges it, one can expect this contribution to be present.

In fact, this term given by the complex solutions of the trajectories appears whenever we have a mixed state. It is natural then that it arises in the torus where we are dealing with a thermal state, as opposed to the case of the plane where we simply have the vacuum. We have also seen that it is present in the circle at zero temperature but only in the Ramond sector. This is because the Ramond sector has a zero mode and then two linearly independent vacuum states appear, which are mixed in the zero temperature state. We may still think of this state as being thermal with arbitrarily large inverse temperature $\beta$ and then use the GNS construction machinery. In any case, this novel term in the modular conjugation is there to remind us that we are dealing with a non-pure state and thus needs to be purified somehow to use the Tomita-Takesaki theory.

\section*{Acknowledgements}

We thank D. Blanco, G. Pérez-Nadal, H. Casini, A. Garbarz and L. Martinek for fruitful discussions. This work has been supported by CONICET.

\appendix

\section{Some properties of $z$}\label{sec:appendixzeta}

In this appendix we prove the bound for the imaginary part of $z$. We will do so invoking the Weierstrass functions and some of its basic properties. Let us first show that $\mathrm{Im}\, z$ is a monotonous function in $\ell$, the length of $V$. That is, we will show that the imaginary part of the integrand in \eqref{z_torus} 
\begin{equation}
     \mathrm{Im} \left[ \zeta(x+iy) - \frac{y}{\beta/2} \zeta(i\beta/2) \right] \equiv I(x+iy) \,,
\end{equation}
satisfies $I(x+iy)<0$ when we restrict to $0 < y < \beta/2$. As a function of $x$ with fixed $y$, we can use the addition theorem for $\wp$ to find extrema of $\mathrm{Im}\, \zeta$ (note that $\wp(x),\wp(iy)\in\mathbb{R}$)
\begin{equation}
    \mathrm{Im}\, \zeta'(x+iy) = \frac{1}{2}\frac{\wp'(x)}{\big[\wp(x)-\wp(iy)\big]^2} \; \mathrm{Im}\, \Big(\wp'(iy)\Big)\,.
\end{equation}
Since $x=L/2$ is a zero of $\wp'(x)$, it must be an extremum of $\mathrm{Im}\, \zeta$. In fact, one can see it is a maximum and, moreover, because of periodicity $x=L/2$ gives its global maximum
\begin{equation}
    \mathrm{Im}\, \zeta(x+iy) \leq \mathrm{Im}\,\zeta(L/2+iy)\,.
\end{equation}
Note that we could have picked $x=0$, but we would have got a minimum instead. Now we show that
\begin{equation}\label{eq:convex}
    \frac{\mathrm{d}^2}{\mathrm{d}y^2} \, I(L/2+iy) = \mathrm{Im}\; \wp'(L/2+iy)\ > 0\,,
\end{equation}
i.e. it is a convex function of $y$: $\wp'(L/2+iy)$ is imaginary and $\wp'(L/2)=\wp'(L/2+i\beta/2)=0$, with no other zeros in the interval. The fact that $\wp'(L/2)$ is a simple zero and $\wp(L/2)$ a minimum implies that $\mathrm{Im}\; \wp'(L/2+iy)$ is positive for $0<y<\beta/2$. 

Finally note that since $\zeta(x)$ is real, $I(x)=0$. Also since $\wp(x+i\beta/2)$ is real, and hence $\mathrm{Im}\,\zeta(x+i\beta/2)$ constant, $I(x+i\beta/2)=0$. A continuous function which is convex and non-vanishing within an interval and approaches 0 at its ends, is necessarily negative. So we have proved our initial claim
\begin{equation}
    \mathrm{Im}\, \left[ \zeta(x+iy) - \frac{y}{\beta/2} \zeta(i\beta/2) \right] \; \leq \; \mathrm{Im}\, \left[ \zeta(L/2+iy) - \frac{y}{\beta/2} \zeta(i\beta/2) \right] \; < \; 0 \,.
\end{equation} 
It is then clear that $\mathrm{Im}\, z$ has its maximum when $V$ is arbitrarily small and its minimum when $V$ is the whole circle. In the first case, trivially $z=i\pi$, while in the second case we readily find (up to an irrelevant real constant) $z(u) = \frac{2 \pi u}{\beta}$. Hence, we conclude
\begin{equation}
    0 < \mathrm{Im}\, z(u) < \pi \,.
\end{equation}
for $0 < \mathrm{Im}\,u < \beta/2$. Finally we remark that from $I(x+i0^+)=-\pi\delta(x)$ and $I(x+i\beta/2)=0$ it follows that $\mathrm{Im}\, z(x+i\beta/2) = \pi$, while $z(x) := \lim_{y \to 0^+} z(x+iy)$ has imaginary part $0$ when $x\in V$ and $\pi$ when $x\notin V$.\\

We have also claimed $\mathrm{Re}\,z(x+i\beta/2)$ is monotonic. We have
\begin{equation}
    z'(x+i\beta/2) = \int_V \mathrm{d}t  \left[ -\wp(x-t+i\beta/2) - \frac{\zeta(i\beta/2)}{i\beta/2} \right]\,.
\end{equation}
For a rectangular lattice the integrand has its minimum when $x-t=L/2$. Then we can expand the Weierstrass functions in terms of the nome $q=e^{-\pi\beta/L}$
\begin{equation}
    -\wp(L/2+i\beta/2) - \frac{\zeta(i\beta/2)}{i\beta/2} \; = \; -\frac{4 \pi ^2}{L^2} \left[ \frac{1}{2 \log q} + \frac{q}{(1+q)^2} - \sum_{n=1}^{\infty} \frac{(-1)^{n} n q^{2 n}}{1-q^{2n}} \left(q^n + q^{-n}\right) \right] > 0 \,.
\end{equation}
We have verified numerically that the expression is indeed positive for all $0<q<1$, implying that the integrand of $z'$ is positive definite and hence $z(x+i\tfrac{\beta}{2})$ is monotonically increasing for all $x$.

\bibliography{biblio}
\bibliographystyle{ieeetr}

\end{document}